\documentclass[12pt]{book}

\usepackage[dvips]{graphicx,color}
\usepackage{makeidx,view}
\makeauthorindex
\makeindex

\begin{document}

\BookTitle{\itshape The Universe Viewed in Gamma-Rays}
\CopyRight{\copyright 2002 by Universal Academy Press, Inc.}
\pagenumbering{arabic}

\chapter{%
Energetic Particles in Clusters of Galaxies and Starburst Galaxies                      
}

\author{%
Heinrich J. V\"olk\\
{\it Max-Planck-Institut f\"ur Kernphysik, D-69117 Heidelberg, Germany}\\
}
\AuthorContents{H.J. V\"olk} 
\AuthorIndex{V\"olk}{H.J.}

\section*{Abstract}
The nonthermal characteristics of starburst galaxies and galaxy clusters are
reviewed and the perspectives for their observation in high energy gamma rays
discussed. It is argued that starburst galaxies should have physically similar and
even simpler nonthermal properties as our Galaxy, the particle transport being
convective in their strong galactic winds. They should be measurable in TeV gamma
rays by large detectors, and in fact the recent detection of NGC~253 is a first
demonstration. Galaxy clusters are equally expected to produce a nonthermal
energetic nuclear particle component that gives unambiguous information about the
entropy production history in clusters. Both types of objects should be main sources
of and at the same time also major visible representatives for the ubiquitous
nonthermal component of matter in the Universe.

\section{Introduction}

In this review I will first discuss the production of energetic particles in starburst
(SB) galaxies and galaxy clusters. The resulting particle density and the energy
spectrum are not determined by the sources alone, but depend to the same extent on the
transport and confinement properties in such objects; especially in clusters of galaxies
the question of confinement plays a decisive role. Therefore it is necessary to address
the combined problem involving both the sources and the propagation (e.g. V\"olk et al.
1996). I shall in fact argue that, at least for strong sources and extended objects,
energetic particle transport corresponds to simple limits. Using multi-wavelength
information from the radio to the X-ray region, as well as recent $\gamma$-ray
observations and theory for Supernova Remnants (SNRs) in our Galaxy, the expected
energetic particle densities are very high, with a Cosmic Ray energy density in SB
galaxies that is about two orders of magnitude higher than in our Galaxy. In clusters,
the average nonthermal energy density corresponds roughly to that in the Interstellar
Medium of the Galaxy. If we consider the enormous gas masses and physical sizes of rich
galaxy clusters, we expect a sizeable overall $\gamma$-ray emission in the sub-TeV and
TeV range which may be observable with the new generation of instruments, both on the
ground and in space, despite the large distances and angular sizes of even nearby
clusters in excess of hundred Mpc.

With this overall picture in mind we can attempt to interpret the observed nonthermal
radiation. For clusters the physical characteristics of the emission in the synchrotron
and Inverse Compton channels depend strongly on the strength of the magnetic field.
Therefore I shall consider separately the cases of 'low' ($\sim 10^{-1}\mu$G) and 'high'
($\sim 1 \, \mu$G)  field strengths, respectively. Even though radio observations indicate
high Intracluster (IC) fields, this may not be true everywhere in a nonuniform IC medium.
Finally, we shall ask ourselves about the perspectives for high energy $\gamma$-ray
astronomy. The result is that nearby SB galaxies should have been marginally observable
with the previous instruments like Whipple, CANGAROO I, HEGRA, or CAT. However, the next
generation of instruments should be able to identify both SB galaxies and clusters of
galaxies as TeV sources, even though, in the case of clusters, not by a large margin.
Although not yet confirmed by another instrument, it is therefore gratifying that the
CANGAROO II telescope appears to have successfully detected the nearby Southern
Hemisphere SB galaxy NGC 253 in the TeV range.

\section{Particle production and propagation in starburst galaxies}
External galaxies should also produce Cosmic Rays (CRs), from sources
similar to those in our Galaxy -- plus possibly through other mechanisms,
like overall accretion or an AGN. SB galaxies are assumed to be analogous
to the Milky Way, although with a significantly enhanced star formation
rate (SFR). This should lead to a higher CR production rate as well.
However, the form and the strength of the galactic mass loss is probably
different. Whereas in our own and similar spiral galaxies, the CR pressure
drives a moderate mass loss (Breitschwerdt et al. 1991; Zirakashvili et
al. 1996)  which ultimately, at large distances, implies convective
removal of all energetic particles produced in the disk (Ptuskin et al.
1997), the wind from SB galaxies is driven by the pressure of an extremely
hot ($\sim 10^8$~K) component of the interstellar gas (e.g. Schaaf et al.  
1989). Like the nuclear CR component, it can basically only cool by
adiabatic expansion, resulting in wind velocities of thousands of km/sec.
Therefore one must assume that beyond the dense (molecular) gas disk, CR
propagation is essentially convective at all particle energies of interest for
$\gamma$-ray astronomy.
\begin{figure}
\parbox{\linewidth}{\centering \includegraphics[width=0.9\linewidth]{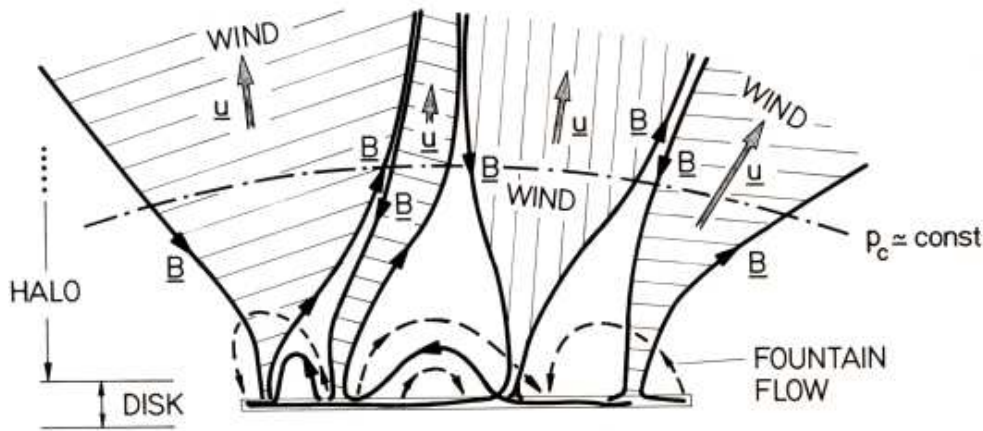}}
\parbox{\linewidth}{\caption{Schematic of galactic wind formation above active 
regions of a galactic disk. Individual lobes of the magnetic field $\underline{B}$ get extended 
into
the halo and beyond with little loss of magnetic flux. They look like funnels (hatched),
while other lobes only delineate fountain flows. At larger distances the flow velocity 
$\underline{u}$
and the gradient of the CR pressure $p_c$ become more or less radial. A central SB nucleus
will produce a strong inner funnel at the expense of others.
}}
\end{figure}
The comparison with the Galaxy is interesting and important for several 
reasons:
\begin{enumerate}
\item To ascertain the contribution of recent star formation to CR
production, compared to other processes (e.g. the action of the central
Black Hole, of SNRs from old stars, and of X-ray binaries from the old
stellar population)
\item To study the CR propagation process, i.e. the relative role of 
convection and diffusion, as indicated above
\item To test the theory of an universal CR origin (Burbidge and Brecher 1971) which
would imply a universal CR intensity.
\end{enumerate}

For definiteness I shall discuss two concrete examples, M82 in the Northern
Hemisphere, and NGC 253, its 'twin' in the Southern Sky. They are the closest SB
galaxies at distances of $\sim 3.2$~Mpc (M 82) and $\sim 2.5$~Mpc (NGC 253).

\subsection{M82} 
M82 is the small sister of the large spiral M81 in the Northern Hemisphere, at a
declination of 70 degrees. The high radio synchrotron intensity in the central
region (galactocentric radius $\simeq 300$~pc; height above disk midplane $\simeq
100$~pc) of M82 suggests very high CR intensities.  From minimum energy arguments
we have a total CR energy density $u_{CR} \simeq 80$~eV/cm$^3$ and a magnetic field
strength $B\simeq 50 \mu$G, assuming an electron to proton ratio of $10^{-2}$,
similar to that in the Galaxy.  Thus, $u_{CR}$ and the magnetic field energy
density $u_B$ may exceed their values in the Galaxy ($\simeq 1$~eV/cm$^3$) by about
two orders of magnitude.

Also the far infrared (FIR) emission is large with an energy density $u_{rad}
\simeq 200\mathrm{eV/cm}^3 \sim u_{rad} \sim u_B$ in the disk, presumably due to
the absorption of the UV radiation from the many young and massive stars by
interstellar dust in the high mean (molecular) gas density $\simeq
140$~H-atoms/cm$^3$ of the SB nucleus. This is consistent with the fact that M82
also obeys the radio/FIR-correlation that characterizes "normal" star forming
galaxies, i.e. galaxies whose emission is dominated by star formation and not by
the radiation from an AGN (e.g. Wunderlich et al. 1987; V\"olk 1989). Observations suggest
that the average magnetic field strength increases about as fast as the radiation
field in starbursts (Lisenfeld et al. 1996).

The extended soft X-ray emission as well as the spectral break in the radio synchrotron
spectrum at a close distance from the CR source region imply a fast, thermally driven
galactic wind with a velocity of several 1000 km/sec (Chevalier and Clegg 1985). As
mentioned before, this strongly suggests that beyond the immediate neighborhood of the
SB nucleus CR transport is dominated by convection in this wind rather than by
diffusion, in contrast to the situation in the lower halo of our Galaxy (Ptuskin et al.
1997). With an observed supernova rate $\nu_{\mathrm{M82}}(SN)\leq 0.3/\mathrm{yr}$,
which is $\leq 10$~times the Galactic SN rate $\nu_{\mathrm{Gal}}(SN)$, it follows that
this convective escape is consistent with SNRs being the sources for the CRs, both for
M82 and the Galaxy, i.e. $dE_{CR}/dt = A \nu(SN)$, where the proportionality constant
$A$ is the same for both galaxies, despite their enormously different characteristics
(V\"olk et al. 1989). The question is then, whether $A$ is a universal constant for
"normal" galaxies. An answer can only be obtained by an investigation of other objects.

\subsubsection{The role of Gamma-ray astronomy}
However plausible such results may be, they can only furnish a necessary
condition for the proof of the supposition that the CR sources are
principally the same in all galaxies. Since the arguments presented depend
on the assumption that the energy density of nuclear CRs exceeds that of
synchrotron radiating CR electrons by a factor of 100, it is necessary to
directly measure the $\pi^0$-decay hadronic $\gamma$-rays in those objects. 

The closest to such a measurement is a detection in TeV $\gamma$-rays which shows at
least the presence of particles in roughly the same energy range, even if a separation
between hadronic and leptonic $\gamma$-ray emission is still required. In fact, the
leptonic Inverse Compton $\gamma$-ray spectrum at very high energies should be softer
than the hadronic spectrum. The two contributions can thus be distinguished by a
broad-band measurement of the $\gamma$-ray energy spectrum. The reason behind this is as
follows: Assuming the CRs to be the result of diffusive shock acceleration in SNRs, then
the CR nuclei will have a spectral index $\alpha^{SNR}_p \simeq 2$. The primary CR
electrons of high energies, on the other hand, will be radiatively cooled inside the SNR
with a corresponding index $\alpha^{SNR}_e \simeq 3$. This leads to an almost identical
photon spectral index $\simeq 1$ for $\pi^0$-decay and Inverse Compton $\gamma$-rays
from the sources themselves (Berezhko et al. 2002; V\"olk et al. 2002). After release
from the sources and upon transport out of the galaxy the electrons will continue to
cool radiatively in the strong radiation field and the interstellar magnetic field, so
that their spatially integrated energy spectrum will have an even softer index
$\alpha^{diff}_e \simeq 4$ leading to a spectral index of their diffuse Inverse Compton
$\gamma$-ray emission $\alpha^{diff}_{IC} \simeq 1.5$. However, the spectral index of
the diffuse hadronic $\gamma$-ray emission will still be $\simeq 1$, assuming strictly
convective CR transport. A second distinguishing feature should be the lower cutoff of
the Inverse Compton $\gamma$-ray spectrum, due to the radiation loss-dominated cutoff of
the electron acceleration process in the sources.

I should add that this last consideration is modified when we go to sufficiently low
electron energies, where radiative cooling inside the SNR sources does not occur during
their life time. This should be the case at $\gamma$-ray energies of some GeV. Below
those energies the diffuse Inverse Compton spectrum should flatten to an index
$\alpha^{diff}_{IC} \simeq 1$, or even somewhat below, making detection with space
instruments at these energies more difficult than one might naively expect.

\subsubsection{Gamma-ray detectability}
What is then the $\gamma$-ray emission from M82 and what are the  
prospects for its detection? 
The SB nucleus has an angular size of about 40 arcsec. Such a region would
clearly be a $\gamma$-ray point source. Even for the high molecular gas
density V\"olk et al. (1996) have estimated a rather low hadronic
$\gamma$-ray flux:
\begin{eqnarray*}
F_{\gamma}(E > 100\, \mathrm{MeV}) & = & 1.3 \times 10^{-9}\, {\rm ph\, cm}^{-2}\, {\rm sec}^{-1} \\
F_{\gamma}(> E) & = & 2.6 \times 10^{-13}(E/\mathrm{1\, {\rm TeV}})^{-1.1}\, {\rm ph\, cm}^{-2}\, {\rm sec}^{-1},
\end{eqnarray*}
where $\alpha^{SNR}_p \simeq 2.1$ has been assumed. A slightly harder 
spectral index $\alpha^{SNR}_p \simeq 2.0$ will not significantly change 
these numbers.
Even though the gas density and the target photon field energy density in
M82 are both about equally strongly increased relative to their Galactic values,
the Inverse Compton flux might be somewhat higher, because its emission
volume should be larger. However M82 has neither been detected by EGRET
nor up to now in TeV $\gamma$-rays, in agreement with the above estimate.
The HEGRA upper limit after 35 hours of observations with the stereoscopic
system corresponds to 2 percent of the Crab flux. However, the future
instruments in the Northern Hemisphere like VERITAS and MAGIC, as well as
GLAST in the GeV range, should have a good chance to detect this galaxy.

\subsection{NGC 253}
NGC 253 at a declination of -25 degrees has a projected optical size of $27.5 \times 6.8
\mathrm{arcmin}^2$. As such it would therefore expected to be an extended $\gamma$-ray
source for the 0.1 degree angular resolution stereoscopic arrays CANGAROO III and
H.E.S.S. in the Southern Hemisphere.
The SB nucleus is very much smaller than this geometrical size of the stellar disk. The
supernova rate is estimated to be between 0.1/yr and 0.3/yr (Bhattacharya et al. 1994,
i.e. at least 3 to 10 times larger than for the Galaxy. The CR energy density $u_{CR}$
exceeds the Galactic value by a factor up to 100. A large radio synchrotron halo rises
to a height of at least an optical disk radius (Carilli et al. 1992), consistent with a
correlation of the soft X-ray with the H$\alpha$ emission over the same region above the
disk (Strickland et al. 2002). The same phenomenon exists for M82 as well. Dynamically
this again implies a thermally driven galactic wind, with an estimated mass loss rate of
about $24 M_{\odot}/yr$ (Cappi et al.  1999). The far infrared (FIR) flux, the gas
density in the SB nucleus, and the magnetic field strength are similar to the values in
M82 (Blom et al. 1999). Indeed, the physical state of the two galaxies appears to be
quite similar, as mentioned above.
Goldshmidt and Rephaeli (1995) have used a specific wind model by Lerche and
Schlickeiser (1980) to devise an Inverse Compton $\gamma$-ray emission model
for this galaxy.
More recent Chandra observations (Weaver et al. 2002) appear to show a heavily
absorbed source of hard X-rays within the nuclear SB region, with an X-ray
luminosity that exceeds the Eddington luminosity of a $1.4 M_{\odot}$ object
within the nuclear SB region.  The ionising source may be an intermediate-mass
Black Hole, similar to a luminous compact X-ray source near the centre of M82
(Matsumoto et al. 2001). The absorbing optically thick molecular torus may
collimate the SB-driven nuclear outflow. If this interpretation is correct,
then these galaxies combine a starburst with AGN activity. As a consequence we
must be careful not to over interprete the $\gamma$-ray results from these
galaxies, because the hard X-ray source might itself be a source of
$\gamma$-rays. Nevertheless, at the moment we will take the point of view that
starbursts can be the dominant effect in many galaxies, like in the ones
considered here, giving us the possibility to learn more about CR production
in the Universe as a result of star formation alone.

\subsubsection{Gamma-ray emission}
Like for M82, EGRET has only obtained an upper limit for the $\gamma$-ray flux above a
few 100 MeV (Blom et al. 1999). The OSSE instrument on CGRO, on the other hand, has
marginally detected $\gamma$-ray emission below 1 MeV (Bhattacharya et al. 1994). Most
recently, the CANGAROO collaboration (Itoh et al. 2002) has announced the detection of
$> 0.5$~TeV $\gamma$-rays with the CANGAROO II telescope at the high significance level
of $\sim 11 \sigma$. If confirmed this is an important result. The interpretation by
Itoh et al. (2002) is in terms of Inverse Compton emission from a slightly extended
source (for an estimated angular resolution of 0.23 degrees). The differential
$\gamma$-ray spectrum was fitted with an expression $\propto E^{-1.5} \exp({-
\sqrt{E}/a})$, quite soft, in order to stay below the EGRET upper limit. This may be
possible, even given the severe radiation losses of the electrons discussed before.

\section{Clusters of galaxies}
Galaxy clusters are the largest gravitationally bound objects in the Universe. In contrast
to individual galaxies, the diffuse Intracluster (IC) gas dominates their visible mass by
factors of a few compared to the stars in the constituent galaxies (e.g. Boehringer 2002).  
Rich clusters, our topic here, contain typically several hundred galaxies. The galaxy
velocities are nevertheless so large that the main gravitational mass of clusters must be
due to a Dark Matter. Due to Big Bang nucleosynthesis constraints this Dark Matter cannot
be of a baryonic nature.

What makes galaxy clusters particularly interesting from a cosmological point of view,
is the belief that they contain a representative sample of matter in the Universe,
confining the primordial mixture during their formation. Typical temperatures of the hot
IC gas amount to several keV, giving it as much or more kinetic and thermal energy as is
contained in the galaxies. This gas is presumably representative of the larger part of
the baryonic Dark Matter in the Universe.

Due to the enormous spatial dimensions of clusters and their turbulent state also the
nonthermal particles should be confined over a Hubble time (V\"olk et al. 1996; see also
Berezinsky et al. 1997). Ultrarelativistic nuclear particles do not suffer any
significant losses during this time either. For this reason we have called the nuclear
nonthermal component in clusters "Cosmological Cosmic Rays", because they have been
stored in the clusters since the beginning of production with the onset of cosmic
structure formation. The mass in energetic particles is of course negligible. On the
other hand, their acceleration is inevitably associated with irreversible processes like
shock dissipation. Whereas the thermal gas can cool radiatively, nuclear CRs do not.
Observations of $\pi^0$-decay $\gamma$-rays allow us therefore to study the entropy
history of cluster formation in an unambiguous way.

From the chemical abundances of the IC gas, whose composition is about 40 percent Solar,
a huge amount of star formation per galaxy must have occurred in evolved clusters like
the Perseus and Coma clusters. This is in line with the dominance of giant elliptical
and lenticular galaxies generally thought to be the result of interactions and mergers
of (proto)spirals. The CR production in the large number of core collapse SNRs
associated with these SBs must have been correspondingly large. The galactic winds
should not only have adiabatically cooled and transported the particles into the IC
medium but should also have strongly reaccelerated them in their termination shocks at
the transition to the IC gas.

If one calculates the energy input into the IC gas from these processes, it follows that
the main energisation of the IC medium, by a factor of more than 10, must have come from
gravitational compression. As shown by V\"olk et al. (1996), this could in principle
have come primarily through adiabatic $pdV$-work at contraction, after the main particle
acceleration events as a result of the SBs, with little further entropy production.
However this is an unlikely extreme, even though the most common shocks in the IC gas
are not very strong, with typical Mach numbers of 4 -- 5 (Miniati et al. 2000).  
Consequently, there is a comparably important possibility of (primary)  particle
acceleration in accretion shocks of subclusters onto the main cluster during ongoing
cluster formation (e.g. Blasi 2002; Miniati 2002).

In addition to these ubiquitous and often large scale acceleration sites also the
energy input from shock heating and acceleration by the large-scale jets from single
FR II galaxies in clusters has been considered (e.g. En\ss lin et al. 1997) and even
argued to be dominant (e.g. Inoue and Sasaki 2001). 

It will be important to distinguish the different processes. We believe that their joint
action will lead to a multi-phase IC medium whose observational study is the key to a
full understanding of the role of the various mechanisms that increase the entropy of
the IC gas.

\subsection{Nonthermal radiation from clusters}
Presumably nonthermal radiation from clusters has been observed as synchrotron emission
at radio wavelengths (e.g. Ferretti and Giovannini 1996), and in the form of extreme UV
(EUV) (e.g. Lieu et al. 1996; Bergh\"ofer et al. 2000) and hard X-ray (Fusco-Femiano et
al. 1999; Rephaeli et al.  1999) 'excesses' (over a single-temperature plasma emission).
No high energy $\gamma$-rays have been detected yet.

The IC magnetic field strengths in clusters are possibly very large, of the order of
$\mu$G.  They could originally be produced by dynamo action in starburst galaxies and
subsequently be extended into the IC medium by galactic winds. The idea is old (Brecher
and Burbidge 1972) and has been taken up independently by V\"olk and Atoyan (2000). At
the present epoch the magnetic fields would be mostly cut off from the galaxies by
magnetic reconnection and be sheared and possibly amplified by the IC flow field (Dolag
et al. 1999).

The strength of the magnetic field is a decisive parameter for the interpretation of the
radio, EUV and hard X-ray emissions. For 'low' fields of the order of $0.1 \, \mu$G, it
is possible to interpret the hard X-ray excess -- and also the EUV excess, observed in
less than a handful clusters -- as Inverse Compton emission by the radio synchrotron
emitting electrons, assumed to be injected steadily into the IC medium for the last
several billion years (Fusco-Femiano et al. 1999; Rephaeli et al. 1999). A break
between the EUV and the hard X-ray regions could then be understood as a result of
radiative losses.

\begin{figure}
\parbox{\linewidth}{\centering \includegraphics[width=0.8\linewidth]{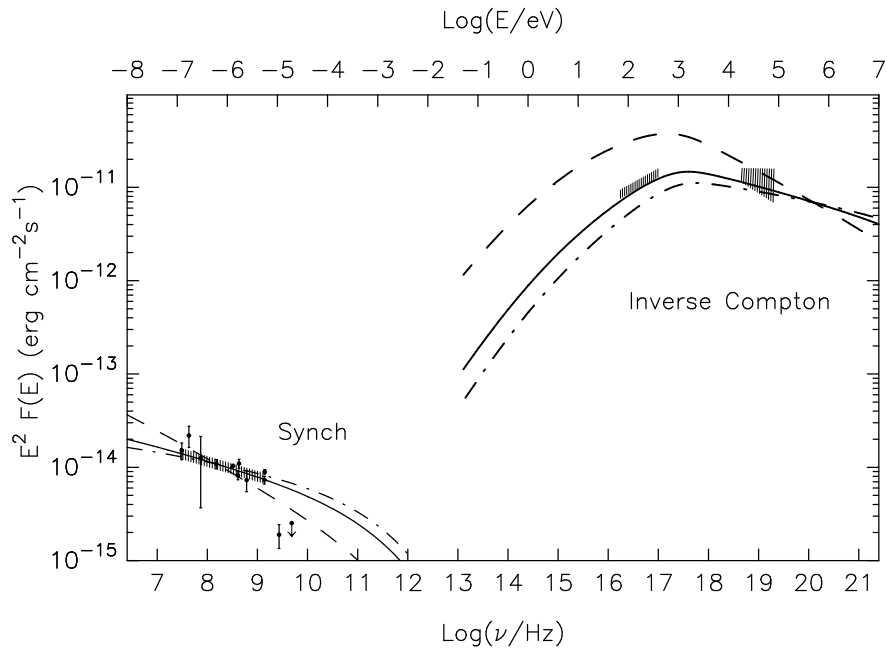}}

\parbox{\linewidth}{\caption{Synchrotron and Inverse Compton fluxes calculated for a 
magnetic field $B = 0.12 \, \mu$G in the Coma cluster, assuming stationary injection
of relativistic electrons with three different power-law energy distributions during
the last $3 \times 10^9$~yr. The radio data are taken from Deiss et al. (1997). The
hatched regions in the EUV and X-ray domains correspond to the fluxes observed by
Lieu et al. (1999), and Rephaeli et al. (1999) and Fusco-Femiano et al. (1999),
respectively. (From Atoyan and V\"olk 2000).} }
\end{figure}

For similar parameters and a mean gas density of $n= 10^{-3}~\mathrm{cm}^{-3}$, the
extension of these spectra into the high energy $\gamma$-ray region is shown in Fig. 3,
also including the electron Bremsstrahlung emission spectrum. There would be a good
chance for GLAST to observe such emission.

\begin{figure}
\parbox{\linewidth}{\centering \includegraphics[width=0.8\linewidth]{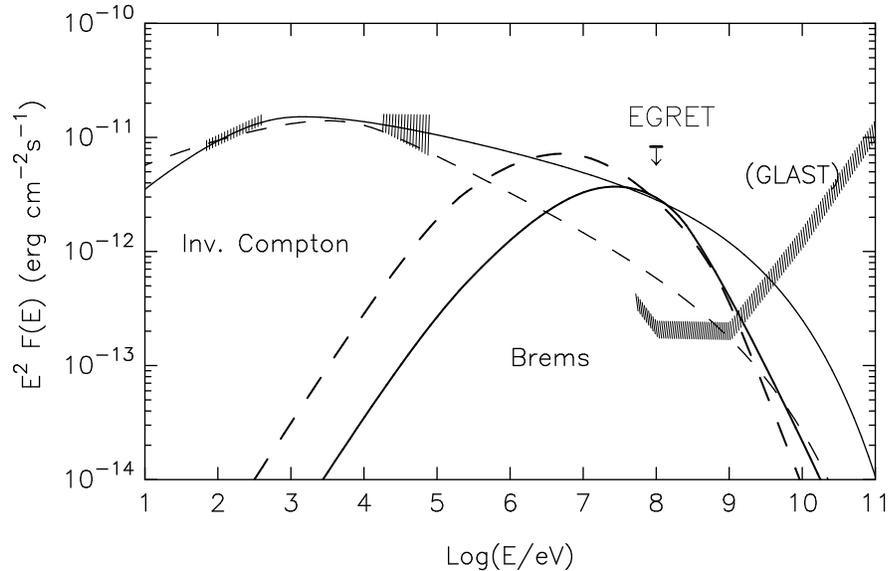}}


\parbox{\linewidth}{\caption{High energy Bremsstrahlung and Inverse Compton
fluxes from the Coma cluster, calculated for similar parameters as in Fig. 2. The
expected flux sensitivity of the GLAST detector (from Bloom 1996), and the upper flux
limit of EGRET (Sreekumar et al. 1996) are also shown. (From Atoyan and V\"olk
2000).} }
\end{figure}

However, this simple interpretation becomes untenable if the magnetic field is
'high', i.e.  of the order of $1 \, \mu$G (Kim et al. 1990; Kronberg 1994), because
then, for given radio intensity, the Inverse Compton emission is very small. For Coma
it is then possible to understand the radio synchrotron emission as the result of a
population of secondary electrons, produced by inelastic collisions of CR protons
with IC gas nuclei and charged pion decay (Blasi and Colafrancesco 1999). Any Inverse
Compton component responsible for the EUV excess must however be cut off before hard
X-ray energies, in order not to overproduce radio synchrotron photons (Fig. 4). This
is possible with a relic (from the early SB history) electron population that
has been radiatively cooled to maximal Lorentz factors of the order of 100, while
being subject to continuing gravitational compression (Sarazin and Lieu 1998). The
hard X-ray excess remains unexplained in this context (Atoyan and V\"olk 2000).

\begin{figure}
\parbox{\linewidth}{\centering \includegraphics[width=0.8\linewidth]{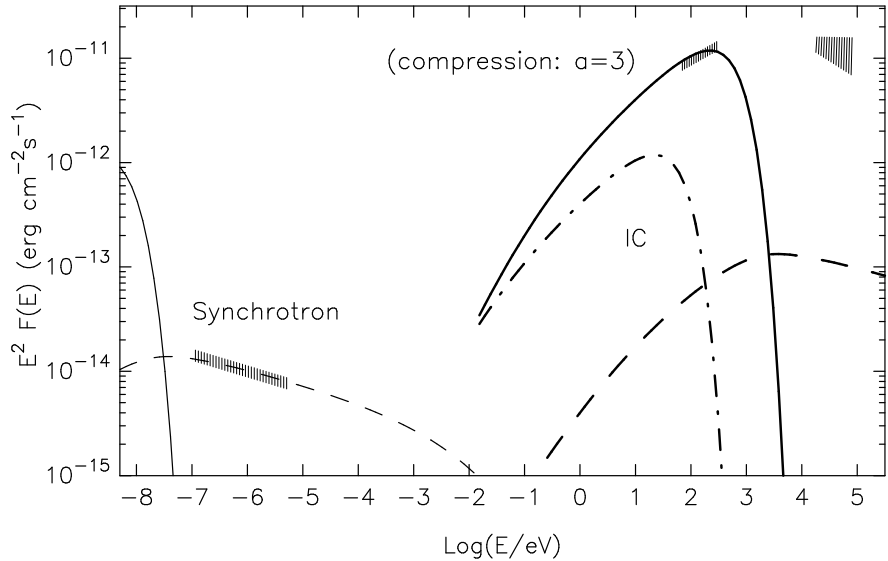}}


\parbox{\linewidth}{\caption{Fluxes of synchrotron (thin lines) and Inverse 
Compton (heavy lines) radiation produced by a relic population of electrons with 
(solid lines) and without (dash-dotted line) later adiabatic compression, and by a 
population of secondary electrons (dashed lines). (From Atoyan and V\"olk
2000).} }
\end{figure}

\subsubsection{Thermal origin of the hard X-ray 'excess'}
This X-ray problem might have a very different solution than by nonthermal
radiation! If we take into account that the SBs, forming the early type galaxies in
the clusters, must have led to strong and fast galactic winds, with speeds of
several 1000 km/sec, similar or in excess to those of present-day SB galaxies such
as M82, then the dissipation in their termination shocks will result in thermal
energies of the wind bubbles capable to produce thermal X-rays in the 10 keV range.
Subsequent gravitational compression gives ample room for adiabatic heating to
temperatures of the order of 100 keV. Such a 'very hot' component of the IC gas
would certainly survive until the present epoch, giving rise to a multi-phase IC
medium in approximate pressure equilibrium. The relatively weak accretion and merger 
shocks can not produce such a component. Its formation would have to be accompanied
by the acceleration of nuclear particles and electrons, so that one must expect
radio synchrotron and thermal hard X-ray emissions to be correlated. Observations of
this kind have been recently reported by Fusco-Femiano et al. (2003). Thus we
suggest that the so-called 'excess' hard X-ray emission is of thermal origin, as a
relic of the early history of galaxy formation in the cluster.

\subsection{High energy $\gamma$-ray emission from clusters; observational 
perspectives}
Assuming 'high' magnetic fields in the cluster, and taking into account that the
acceleration processes in the IC gas should produce a hadronic CR component as well,
we can estimate the $\gamma$-ray emission. We shall do this for the case of the Coma
cluster.

\begin{figure} 
\parbox{\linewidth}{\centering \includegraphics[width=0.8\linewidth]{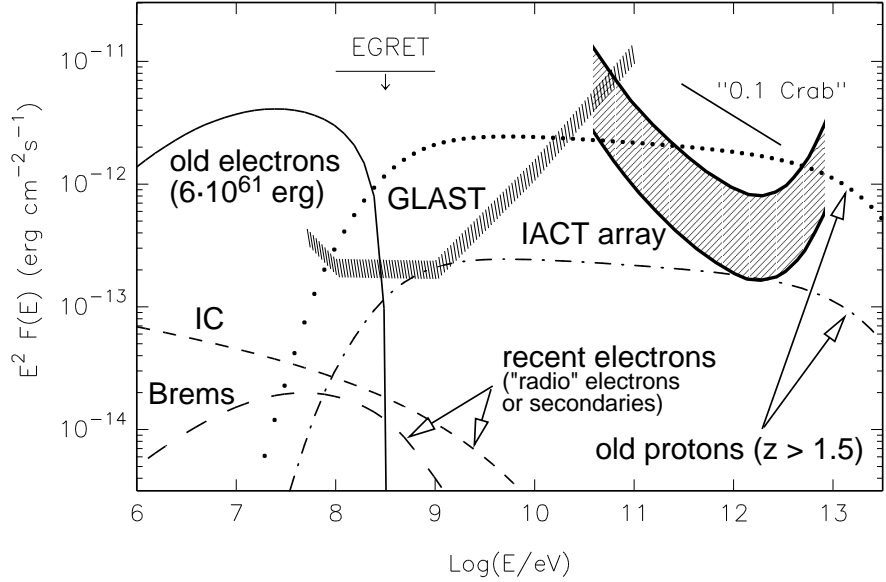}}

\parbox{\linewidth}{\caption{Expected $\gamma$-ray fluxes from the Coma cluster
for $B = 1 \, \mu$G. The solid curve shows the Bremsstrahlung flux produced by the
relic population of old electrons that can explain the EUV 'excess' in Fig. 4. Heavy
and light dashed curves show the Bremsstrahlung and Inverse Compton fluxes due to
the recent secondary radio electrons, respectively. The hadronic $\gamma$-ray   
emission due to nuclear energetic particles (protons) is given for the two levels of
total energy $W_p = 3 \times 10^{62}$ and $3 \times 10^{61}$~ergs by the dotted and 
dash-dotted lines, respectively. The sensitivity of the GLAST detector is given by
the hatched curve, ending at $10^{11}$~eV, together with that of an array of imaging
atmospheric Cherenkov telescopes (IACTs) like CANGAROO III, H.E.S.S., and VERITAS
for a point source (lower thick solid curve) and an extended source (upper thick
solid curve), and contrasted with a spectrum corresponding to 10 percent of the Crab
Nebula flux. Population of secondary electrons (dashed lines). (Based on Atoyan and
V\"olk 2000).} } \end{figure}

The relic electron component whose Inverse Compton emission is assumed to be
responsible for the observed EUV 'excess' is required to have an energy $W_e \sim
5\times 10^{61}$~erg presently. For a mean density $N_{gas} \simeq
10^{-3}~\mathrm{cm}^{-3}$ of the IC gas in Coma, this implies a large Bremsstrahlung
flux up to about 200 MeV (Fig. 5). On the other hand, the Inverse Compton and
Bremsstrahlung $\gamma$-ray fluxes produced by the recent secondary radio electrons
are negligible. A total energy $W_p \leq 3 \times 10^{62}$~ergs in energetic protons
is required to produce these secondaries. This is less than 10 percent of the total
thermal energy in the IC gas, and certainly about what one should expect for
diffusive acceleration in the shocks heating the gas. These protons could indeed be
'old', or 'cosmological', since they do not cool or escape during a Hubble time. The
hadronic $\gamma$-rays should dominate above the sharp Bremsstrahlung cutoff with a
hard differential spectrum that should have a power law index of about 2. Although
below the EGRET upper limit, the Bremsstrahlung and $\pi^0$-decay $\gamma$-rays
should be well within the reach of the future GLAST detector in the GeV range. At
high energies, where no electronic emission can occur any more, ground-based arrays
of imaging atmospheric Cherenkov telescopes like CANGAROO III, H.E.S.S., or VERITAS
should be marginally able to detect objects like the Coma cluster in the TeV range,
even though it is an extended source, with a radius in excess of 1 degree and with a
flux level at 1 TeV of about 5 percent of the Crab Nebula.

%
\section{References}
\noindent
Berezhko, E.G. et al., A\&A, 395, 943 (2002)\\
Berezinsky, V.S. et al., ApJ, 487, 529 (1997)\\ 
Bergh\"ofer, T. et al., ApJ, 535, 615 (2000)\\
Bhattacharya, D. et al., ApJ, 437, 173 (1994)\\
Blasi, P., astro-ph/0207361 (2002)\\
Balsi, P. and Colafrancesco, S., Astropart. Phys., 12, 169 (1999)\\
Blom, J.J. et al., ApJ, 516, 744 (1999)\\
Boehringer, H., Space Sci. Rev., 100, 49 (2002)\\
Brecher, K. and Burbidge, G.R., ApJ, 174, 253 (1972)\\
Breitschwerdt, D. et al., A\&A, 245, 79 (1991)\\
Burbidge, G.R. and Brecher, K., Comments Astrophys. Space Phys., 3, 140 (1971)\\
Cappi, M. et al., A\&A, 350, 777 (1999)\\
Carilli, C.L. et al., ApJ, 399, L59 (1992)\\
Chevalier, R.A. and Clegg, A.W., Nature, 317, 44 (1985)\\
Deiss, B.M. et al., A\&A, 321, 55 (1997)\\
Dolag, K.M. et al., A\&A, 348, 351 (1999)\\
En\ss lin, Th. et al., ApJ, 477, 560 (1997)\\
Fusco-Femiano, R. et al., ApJ, 513, L21 (1999)\\
Fusco-Femiano, R. et al., A\&A 398, 441 (2003)\\
Goldshmidt, O. and Rephaeli, Y., ApJ, 444, 113 (1995)\\
Inoue, S. and Sasaki, S., ApJ, 562, 618 (2001)\\
Itoh, C. et al., A\&A, 396, L1 (2002)\\
Kim, K.-T. et al., ApJ, 355, 29 (1990)\\
Kronberg, P.P., Rep. Prog. Phys., 57, 325 (1994)\\
Lerche , I. and Schlickeiser, R., ApJ, 239, 1089\\
Lieu, R. et al, Science, 274, 1335 (1996)\\
Lisenfeld, U. et al., A\&A, 314, 745 (1996)\\
Matsumoto, H. et al., ApJ, 547, L25 (2001)\\
Miniati, F. et al., ApJ, 542, 608 (2000)\\
Miniati, F., astro-ph/0206430\\
Ptuskin, V.L. et al., A\&A, 321, 434 (1997)\\
Rephaeli, Y. et al., ApJ 511, L21 (1999)\\
Sarazin, C.L. and Lieu, R., ApJ, 494, 177 (1998)\\
Schaaf, R.. et al., A\&A 336, 722 (1989)\\
Sreekumar, P. et al. ApJ, 464, 628\\
Strickland, D.K. et al., ApJ, 565, 689 (2002)\\
V\"olk, H.J., A\&A, 218, 67  (1989)\\
V\"olk, H.J. et al., A\&A, 213, L12 (1989)\\
V\"olk, H.J. et al., Space Sci. Rev., 75, 279 (1996)\\
V\"olk, H.J. et al., A\&A, 396, 649 (2002)\\
Weaver, K.A. et al., ApJ, 576, L19 (2002)\\
Wunderlich, E. et al., A\&AS, 69, 487 (1987)\\
Zirakashvili, V.L. et al., A\&A, 311, 113 (1996)\\

\endofpaper
\end{document}